\documentclass[USenglish,twocolumn]{article}

\usepackage[utf8]{inputenc}%(only for the pdftex engine)
\usepackage[big]{dgruyter}

\begin{document}

%  \articletype{Research Article{\hfill}Open Access}

  \author*[1]{Francesco Ciccarello}

%\author[2]{}

  \affil[1]{NEST, Istituto Nanoscienze-CNR and Dipartimento di Fisica e Chimica, Universit$\grave{a}$  degli Studi di Palermo, via Archirafi 36, I-90123 Palermo, Italy, E-mail: francesco.ciccarello@unipa.it}

%  \affil[2]{Affil, E-mail: @email.edu}

  \title{\huge Collision models in quantum optics}

  \runningtitle{Collision models in quantum optics}

  %\subtitle{...}\newcommand{\eq}{Eq.~}
  \newcommand{\eq}{Eq.~}
  \newcommand{\eqs}{Eqs.~}
  \newcommand{\fig}{Fig.~}
  \newcommand{\figs}{Figs.~}
  \newcommand{\cf} {cf.~}
  \newcommand{\ug} {\!=\!}
  \newcommand{\tens} {\!\otimes\!}
  \newcommand{\piu} {\!+\!}
  \newcommand{\meno} {\!-\!}
  \newcommand{\ie} {i.e.~}
  \newcommand{\eg} {e.g.~}
  \newcommand{\av}[1]{\langle#1\rangle}
  \newcommand{\up}{\uparrow}
  \newcommand{\dow}{\downarrow}
  \newcommand{\rref} {Ref.~}
  \newcommand{\rrefs} {Refs.~}
  \newcommand{\cia} {IOT }

\begin{abstract}
{Quantum collision models (CMs) provide advantageous case studies for investigating major issues in open quantum systems theory, and especially 
	quantum non-Markovianity. After reviewing their general definition and distinctive features, we illustrate the emergence of a CM in a familiar quantum optics scenario. 
	This task is carried out by highlighting the close connection between the well-known input-output formalism and CMs. Within this quantum optics framework, 
	usual assumptions in the CMs' literature -- such as considering a bath of non-interacting yet initially correlated ancillas -- have a clear physical origin.}
\end{abstract}

  \keywords{collision models; quantum non-Markovian dynamics; input-output formalism}
%  \classification[PACS]{}
 % \communicated{...}
 % \dedication{...}

  \journalname{}

%\DOI{DOI}
  %\startpage{1}
%  \received{..}
%  \revised{..}
%  \accepted{..}

%  \journalyear{2014}
%  \journalvolume{1}
%  \journalissue{11}

\maketitle
\section{Introduction}

The effective description of the dynamics of an open quantum system, \ie one in contact with an external environment, is arguably one of the most daunting problems in quantum mechanics. No 
general equation governing such non-unitary dynamics is known except in few special cases, the most prominent and conceptually important being a Markovian dynamics for which the celebrated Gorini–Kossakowski–Sudarshan–Lindblad master equation (ME), or Lindblad ME in short, is the widespread descriptive tool \cite{books}. The purpose of attacking {\it non-Markovian} (NM) dynamics is yet currently strengthening \cite{reviewNMM,review-vega}, which in particular calls for a deeper understanding of the mechanisms causing quantum NM behaviour. Along this line, an emerging approach is to use quantum collision models (CMs) or, better to say, NM generalisations of CMs \cite{ciccarello2013a,ciccarello2013b,mccloskey2014,sciarrino2015a,lorenzo2016,luoma2016,pezzuto2017,giovannetti2012a,giovannetti2012b,cusu1,cusu2,
	lorenzo2017,grimsmo2015,whalen2017,rybar2012,santos2014,sciarrino2015b,santos2016,santos2017,vegaCM,sabrinaCM,dariusCM,rodriguez2008,aspuru2012,rodriguez2011}. 
The basic version of a CM \cite{rau,alicki1987,brun,scarani2002,ziman2002,ziman2003,bruneau2014b} 
considers a system $S$ in contact with a bath $B$, the latter being made up of a large number of smaller non-interacting particles or ``ancillas". The dynamics proceeds through successive pairwise ``collisions" between $S$ and the bath ancillas, each collision being typically modeled as 
a unitary operation on $S$ and the involved ancilla. If the ancillas are initially uncorrelated (bath in a {\it product} state) and each of them collides with $S$ only once, such a model -- in fact by contruction -- leads to a Markovian dynamics for $S$ which in the continuous-time limit is governed {\it exactly} by a Lindblad ME \cite{brun,buzek2005}. The last property alongside
their simple and intrinsically discrete nature make CMs advantageous case studies to investigate major open problems in quantum non-Markovianity once the basic model 
outlined above is modified so as to introduce a memory mechanism. Among the ways to endow a CM with memory are: adding ancilla-ancilla collisions \cite{ciccarello2013a,ciccarello2013b,mccloskey2014,sciarrino2015a,lorenzo2016,luoma2016,pezzuto2017}, embedding $S$ into a larger system \cite{giovannetti2012a,giovannetti2012b,cusu1,cusu2,lorenzo2017},  allowing $S$ to collide with each ancilla more than once \cite{grimsmo2015,whalen2017}, assuming a {\it correlated} initial bath state instead of a product one \cite{rybar2012,santos2014,sciarrino2015b,santos2016,santos2017,vegaCM,sabrinaCM,dariusCM} or initial system-bath correlations  \cite{rodriguez2008,aspuru2012,rodriguez2011}. 
Typical tasks that can be accomplished through NM CMs constructed in one of these ways are: deriving 
well-defined (i.e., unconditionally completely positive) NM MEs \cite{ciccarello2013a,ciccarello2013b, vacchini,darius,piecewise}, gaining quantitative information about the role of system-bath 
and/or intra-bath correlations in making a dynamics NM \cite{mccloskey2014, pezzuto2017, santos2014,sciarrino2015b,santos2016,santos2017}, simulating highly NM dynamics or indivisible channels \cite{rybar2012,sciarrino2015a,sabrinaCM}.

A beginner who first approaches CMs might be naturally concerned with the predictive power of these models with respect to really occurring open dynamics \cite{nota-nature}. 
Concerns may arise such as the
following ones. Since $S$ interacts with one bath ancilla at a time, the interaction Hamiltonian between $S$ and the bath (i.e., all the ancillas) is necessarily time-dependent. Thereby, despite its microscopic nature, a CM in fact assumes a {\it time-dependent} system-bath Hamiltonian. This may appear weird since one expects a microscopic environmental model to treat $S$ and $B$ jointly as a closed system. Furthermore, most CMs assume no internal bath dynamics, which can again look unnatural for a number of reasons. One of these regards CMs where ancillas are assumed to be non-interacting with each other but initially correlated: how can bath ancillas happen to be correlated, or even be in a pure entangled state, if no coupling between them is assumed? 

A possible reply to such questions is that CMs should be intended as theoretical toy models enabling to address conceptual issues in open quantum systems theory, which would be most probably intractable with standard system-bath microscopic models. Still, it could be objected that, in order to be useful, the knowledge acquired within a CM framework should eventually be  
translated anyway into real open dynamics.

In this paper, mostly motivated by the need for lessening the seemingly abstract nature of CMs, we consider a typical quantum optics setup described by a usual time-independent system-bath
Hamiltonian and highlight how one can construct a discrete CM which in the continuous-time limit fully reproduces the dynamics. The setup comprises an unspecified system $S$, which in practical cases will consist of one or more atoms and/or cavity modes, that is coupled to a white-noise bosonic bath. As is usual in quantum optics, these situations can be described through the powerful 
{input-output} formalism \cite{gardiner}. We will illustrate how the essential idea behind input-output formalism is in fact the same as the one underpinning a CM. The known time-discretisation procedure of the dynamical evolution in the {input-output} formalism, an approach that is becoming more and more adopted these days \cite{genoni,pichler,whalen2017,zoller,vuckovic,milburn2017,SHL} (\eg in connection with weak continuous measurements), indeed can be seen as the definition of a discrete CM.

Within this framework, the apparently abstract CM assumptions mentioned above become natural and physically clear. Moreover, it is 
clarified the physical origin of an attractive feature of CMs, namely the fact that (if no memory mechanisms are introduced) the Lindblad ME can be worked out with no approximations. In addition, we wil see that the quantum optics framework provides paradigmatic dynamics that effectively illustrate the generally delicate passage to the continuous-time limit that one usually carries out in CMs, in particular the necessity of involving even the ancilla's state in the limit.

As mentioned earlier, the relationship between CMs and input-output formalism, which is our focus here, is somehow implicit in a number of quantum optics works.
Still, to our knowledge, this connection was not made explicit in the Physics literature especially from the viewpoint of open quantum systems theory \cite{attal}. It is
significant in this respect that in a very recent broad review on quantum NM dynamics \cite{review-vega} both input-output formalism and CMs are featured topics but not related to each other. 
Highlighting this link explicitely is the main purpose of the present work.

This paper could also be viewed as a friendly, brief introduction to quantum CMs, where the CM constructed
in the quantum optics scenario works an effective, specific illustration of the general theory. 

We start in Section \ref{defCM} by reviewing some basics of CMs, in particular the passage
to the continuous-time limit and the derivation of the Lindblad ME. In Section 3, after reviewing 
the input-output formalism, we show how the time-discretisation procedure defines a CM which,
depending on the field's initial state, can lead to a Lindblad ME. The specific form taken by this,
in particular whether or not the system Hamiltonian is modified, again depends on the field initial state. This is
shown explicitly in the paradigmatic cases of the vacuum state and a coherent state. We finally spotlight
how, in the general case, the constructed CM generally features a bath that is initially in a correlated state.
In Section 4, we illustrate how the quantum optics framework clarifies the reasons why
CMs lead to Lindblad MEs with no need for approximations. Finally, in Section 5, we draw our conclusions.

\section{Collision models}\label{defCM}

Consider a quantum system $S$ is in contact with a bath $B$. The  bath  is assumed to be a large collection of smaller constituents, or `ancillas', $\{B_n\}$, which are supposed to be all identical and {\it non-interacting} with each other. The Hilbert-space of both $S$ and $B_n $can be of any dimension. It is assumed that the initial $S$-$B$ joint 
state is 
\begin{equation}
\sigma_{0}\ug\rho_{0}\otimes\left(\eta\otimes\eta\otimes...\right)\,,\label{sigma0}
\end{equation}
where $\rho_0$ is the initial state of $S$, while $\eta$ is the initial state common to all the ancillas (tensor product symbols are most of the times omitted henceforth). Note that the initial state of $B$ is a {\it product} state, i.e., the ancillas are initially uncorrelated.

The dynamics is assumed to take place through successive ``collisions", namely pairwise short interactions, between $S$  and each reservoir ancilla: $S$-$B_1$, $S$-$B_2$, $S$-$B_3$,...in a way that at each step $S$  collides with a ``fresh" ancilla that is still in state $\eta$ (each ancilla collides with $S$ only once). A sketch of the collision dynamics is shown in \fig1(a).
 \begin{figure}
	\includegraphics[width=7.5cm]{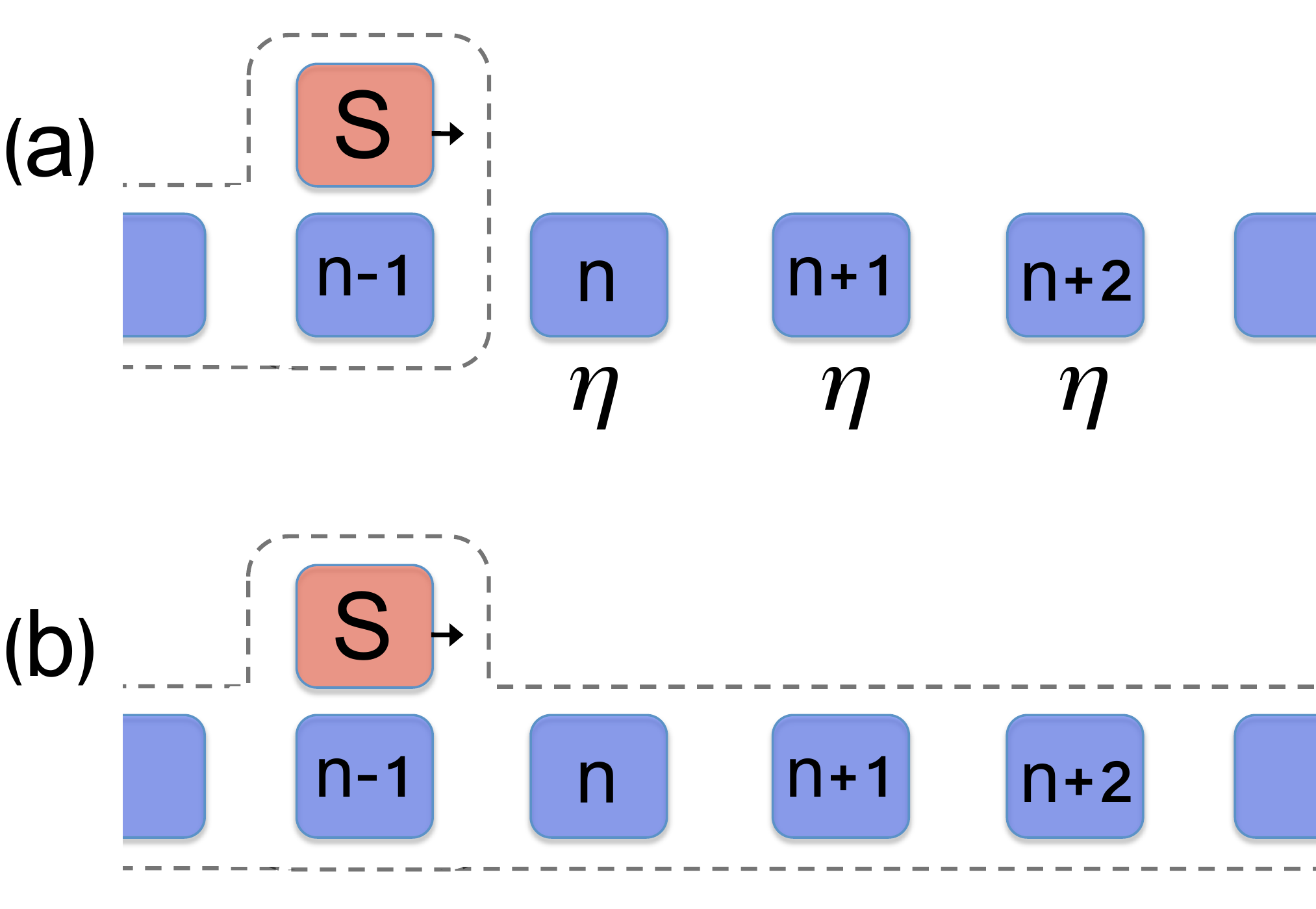}
	\caption{(a) Basic collision model with initially uncorrelated ancillas. The system $S$ collides successively with the bath ancillas, each being initially in state $\eta$ according to \eq\eqref{sigma0}. The figure shows the end of step $n{-1}$, right before the $S$-$n$ collision starts: the system is entangled with all the bath ancillas up to the $(n{-}1)$th but disentangled with the $n$th one. (b) Initially correlated ancillas. The system is generally entangled with all the bath ancillas, in particular with the $n$th one even before the $S$-$n$ collision starts.  \label{fig1}}
\end{figure}

It is assumed that all the collisions have the same duration $\Delta t $, each being described by the unitary evolution operator $\hat U_{n}$ on $S$ and $B_n$ given by (we set $\hbar\ug1$ throughout)
\begin{equation}
\hat {U}_{n}=e^{-i\left(\hat H_{S}+\hat{V}_{Sn}\right) \Delta t }\,,\label{USn1}
\end{equation} 
with 
\begin{equation}
\hat H_{S}=\omega_0 \hat h_S\,,\,\,\,\hat V_{Sn}=g \,\hat v_{Sn}\label{Hs}
\end{equation}
being respectively the free Hamiltonian of $S$ and the interaction Hamiltonian between $S$ and $B_n$. Here, $\omega_0$ and $g$ are the characteristic frequencies  of 
$\hat H_S$ and $\hat v_{Sn}$, respectively, while $\hat h_{S}$ and $\hat v_{Sn}$ are dimensionless operators \cite{nota-rates}. It is assumed
that $B_n$ has no internal dynamics or, alternatively, that the present dynamics is the one occurring in the interaction picture with respect to $\hat H_0=\sum_n\hat H_n$ with $\hat H_n$ 
the free Hamiltonian of the $n$th ancilla \cite{nota-int}.

After $n$ collisions, the overall system is in state
$$\sigma_n=\hat U_{n} \cdots\hat{ U}_{1}\,\sigma_{0}\,\hat U_{1}^\dag\cdots\hat{ U}_{n}^\dag\,.$$
The corresponding state of $S$ is obtained through a partial trace over $B$ as
\begin{eqnarray}
\rho_{n}{=}{\rm Tr}_{B} \{\sigma_n\}=
{\rm Tr}_{B_n}\!\left\{\hat {U}_{ n }\left(\rho_{n-1}\,\eta\right)\,\hat {U}_{ n }^\dag\right\}\,.\label{rhon1}
\end{eqnarray} 
The last identity follows from the fact that -- so long as it has not collided with 
$S$ -- each ancilla remains in the initial state $\eta$ [\cf\eq\eqref{sigma0}] and, most importantly, is fully uncorrelated with $S$. 
This is a distinctive feature of the CM, following in particular from the absence of direct interactions between the ancillas 
and the hypothesis of uncorrelated initial state \eq\eqref{sigma0}. Identity \eqref{rhon1} can be expressed in terms
of the completely positive \cite{books} quantum map 
\begin{eqnarray}
{\cal E}[\rho]={\rm Tr}_{B_n}\!\left\{\hat {U}_ {n} \left(\rho\,\,\eta\right)\,\hat {U}_{ {n} }^\dag\right\} \label{eps}
\end{eqnarray}
as
\begin{equation}
\rho_n=\mathcal E [\rho_{n-1}]\,.\label{rhon2}
\end{equation}
Map $\cal E$ is $n$-independent since $\hat{ U}_{n}$ is assumed to be formally the same for all the ancillas and each of these 
is initially in the same state $\eta$. It follows from \eq\eqref{rhon2} that
$\rho_n={\cal E}^n[\rho_{0}]$, i.e., the evolution of $S$ occurs through iterated applications of $\cal E$ on the initial state $\rho_0$.
\eq\eqref{rhon2} shows that the open dynamics of $S$ is manifestly Markovian (according to any non-Markovianity measure \cite{reviewNMM}) since the evolution of $S$ 
at all steps $n'\ge n$ depends only on the state of $S$ at step $n$: the system keeps no memory of its past history. In more rigorous terms, 
\eq\eqref{rhon2} entails that the (discrete) dynamical map of $S$ is given by 
\begin{equation}
\Phi_n={\cal E}^n \label{DM}
\end{equation}
and thus fulfills the discrete version of the well-known semigroup property \cite{books}
\begin{equation}
\Phi_{n}=\Phi_{n-m} \,\Phi_m\,\label{SG}
\end{equation}
for any $0 \le m \le n$.
Continuous-time dynamical maps fulfilling the semigroup property are well-known to be governed by the Lindblad ME \cite{books}. 
The continuous-time limit of a CM is thus expected to yield a 
Lindblad ME governing the dynamics of $S$, as we show next.

\subsection{Continuous-time limit}

\subsubsection{Change of $\rho_{n}$ per unit step}

In the above discussion, the duration of each collision $\Delta t$ could be any. To pass to the continuous-time limit, we require $\Delta t$ to be small enough 
in a way that $\hat {U}_{n}$ [\cf\eq(\ref{USn1})] can be approximated as
\begin{equation}
\hat {U}_{n }\simeq \hat{\mathbb I}-i (\hat{H}_{S}+{\hat V_{Sn} })\Delta t -\frac{{\hat V}_{Sn}^2}{2}\Delta t ^2\,,\label{USn2}
\end{equation} 
where $\hat{\mathbb I}$ is the identity operator.
Note this is a second-order approximation with respect to $\hat V_{Sn}$ but of the first order in $\hat H_S$; 
hence in \eq\eqref{USn2} it is implicitly assumed that [\cf\eq\eqref{Hs}]
\begin{equation}
\omega_0\ll g \,.\label{approx1}
\end{equation} 
Based on \eq\eqref{rhon2}, the change of the state of $S$ per unit step reads
\begin{equation}
\Delta \rho_n=\rho_{n+1}{-}\rho_{n}=\left(\mathcal{E}-\mathcal{I}\right)[\rho_{n}]\,
\end{equation}
with $\mathcal I$ the identity map.
By replacing next the expression taken by map ${\cal E}$ [\cf\eq\eqref{eps}] when $\hat U_n $ is approximated as in \eq\eqref{USn2}, we get
 \begin{align}
{\Delta \rho_n}&=-i \,[\hat H_S,\rho_{n}]\Delta t-i\,{\rm Tr}_{B_n}\!\left\{[\hat{V}_{\mathcal {S}n},\rho_n\eta]\right\}\!\Delta t\nonumber\\
&+\, {\rm Tr}_{B_n}\!\left\{{\hat V}_{Sn}(\rho_n\eta) \,{\hat V}_{Sn}\!
\meno\frac{1}{2}\left[{\hat V}_{Sn}^2,\rho_n\eta\right]_+\right\}\!\Delta t^2\label{drho}\,
\end{align}
with $[\cdots,\cdots]_+$ denoting the anticommutator and where we dropped third-order terms in $\Delta t$. 
Now, we note
that the two partial traces in the above equation define, respectively, an effective Hamiltonian and Lindblad dissipator both acting on $S$ according to
\begin{eqnarray}
\hat H'_{S}&=&{\rm Tr}_{B_n}\{\hat V_{Sn}\eta\}=g\,{\rm Tr}_{B_n}\{\hat v_{Sn}\eta\}\,, \label{HpS}\\
{\mathcal D}[\rho]&=&\Gamma\,\,\sum_{ij} \left(\hat{L}_{ij}\rho\hat{L}_{ij}^\dag\meno\frac{1}{2}[\hat L_{ij}^\dag\hat L_{ij},\rho]_+\right)\,,\label{diss}
\end{eqnarray}
with
\begin{equation}
\Gamma=g^2\Delta t\,\label{rate}
\end{equation}
and the jump operators $\{\hat{L}_{ij}\}$ defined by
\begin{equation}\label{jump} 
\hat{L}_{ij}=\sqrt{p_j}\,\,%_{B_n}
\!\langle i|\hat{v}_{ {S}n}|j\rangle\,,
\end{equation}
where we used \eq\eqref{Hs}. Here, probabilities $\{p_j\}$ come from the 
eigenstate decomposition of the initial ancilla's state $\eta=\sum_k\,p_k\,|k\rangle\langle k|$
with $\{|i\rangle\,,j\rangle\}$ standing for a pair of kets taken from the orthonormal basis $\{|k\rangle\}$
and with the sum in \eq\eqref{diss} running over all possible pairs (see also \rref\cite{lorenzo2017}).
Note we absorbed rate \eqref{rate} in the definition \eqref{diss} in such a way that $\cal D$ has
dimensions of a frequency (just like $\hat H_S'$). 

By replacing \eqs\eqref{HpS} and \eqref{diss} into \eq\eqref{drho} and dividing each side by $\Delta t$,
the change of the state of $S$ per unit step reads
 \begin{align}
\frac{\Delta \rho_n}{\Delta t}&=-i \,[\hat H_S+\hat H'_S,\rho_{n}]+ \,{\cal D}[\rho_n]\label{drho2}\,.\,
\end{align}

\subsubsection{Lindblad master equation}\label{LME}

It should be already clear to many readers that \eq(\ref{drho2}) is in fact a discrete version of the Lindblad ME.
The task is now to work out a standard Lindblad ME where time is a continuous variable as usual, this being a 
step in our treatment that needs some care. Let us assume first that we want to describe the dynamics up
to time $t=N \Delta t$, where $N$ is the total number of collisions. Hence,
\begin{equation}
\Delta t=\frac{t}{N} \,\,.\label{step}
\end{equation}
Correspondingly, we define 
a discrete time variable as $t_n=n \Delta t$ with $0\le n\le N$. 
As is usual when performing continuous limits we fix $t$ (which can be arbitrary though) and let $N\rightarrow \infty$. 
Thereby, $\Delta t\rightarrow 0$ according to \eq\eqref{step}. 

 So far, we implicitly treated the model parameters as fixed constants. It is clear however
 that if this were the case then -- assuming that the ancilla's state $\eta$ is kept fixed -- the rate $\Gamma$ [see \eq\eqref{rate}] would vanish as $\Delta t\rightarrow 0$ and the dynamics of $S$ would be unitary
 with \eq\eqref{drho2} reducing to a Von Neumann equation with Hamiltonian $\hat H_S+ \hat H'_S$. This is neither awkward nor trivial, 
 as recently highlighted in \rrefs\cite{altamirano2017a,kempf2016}, and can give rise to appealing applications such as the implementation of one- and two-qubit quantum gates \cite{datta}.
In order for the dissipator in \eq\eqref{drho2} to survive in the $N\rightarrow\infty$ limit when $\eta$ is kept fixed, we necessarily need to demand 
$g$ to grow with $N$ in such a way that rate \eqref{rate} converges to a finite value for $N\rightarrow \infty$. Yet, this can raise concerns since if $g\rightarrow \infty$ 
then one might expect $\hat H'_S$ to diverge [\cf\eq(\ref{HpS})]. This issue is typically got around by assuming that the average in \eq(\ref{HpS}) is zero -- which is true
in many typical situations -- or by invoking a renormalization of the free Hamiltonian of $S$. Later on (see Subection \ref{coherent}) we will see that $\hat H'_S$ can happen to survive
the continuous-time limit (alongside the dissipator $\cal D$) due to the fact that even the ancilla's state $\eta$ must in general be regarded as $N$-dependent and 
therefore involved in the limit.

Based on the above discussion, we conclude that in the continuous-time limit \eq\eqref{drho2} is turned into
the ME
 \begin{align}
\frac{{\rm d}\rho}{{\rm d} t}=-i \,[\hat H_S+\hat H'_S,\rho]+ \,{\cal D}[\rho]\label{ME}\,\,
\end{align}
with $\hat H'_S$ and ${\cal D}$ given by the $N\rightarrow \infty$ limit of \eqs\eqref{HpS} and \eqref{diss}, respectively.
A more rigorous and complete discussion on the derivation of the continuous-time ME can be found in \rref\cite{altamirano2017a}.

As expected from the intrinsically memoryless nature of the CM, which was highlighted at the end of the previous section, 
\eq\eqref{ME} is a Lindblad ME. Thereby, the basic version of CM presented here defines a fully Markovian dynamics in the continuous-time
limit.

\subsubsection{Time-dependent Lindblad ME}\label{TDLME}

In the above, for the sake of argument we considered a fully homogeneous CM. One can straightforwardly 
generalise the above treatment to the case that the system free Hamiltonian $\hat H_S$, the ancilla's state $\eta$ and the interaction Hamiltonian $\hat V_{Sn}$
are all dependent on the step number $n$. Accordingly, the completely positive quantum map \eqref{eps} describing the system's evolution in 
a single collision will become $n$-dependent as well, i.e., ${\cal E}\rightarrow{\cal E}_n$, with the discrete dynamical map \eqref{DM} now
generalised as
$$\Phi_n={\cal E}_n\,\cdots\,{\cal E}_1\,.$$
This no longer obeys the standard semigroup property \eqref{SG}. Yet, the above equation shows that the dynamics
 can be divided into a succession of completely positive (CP) quantum maps, each of which will become an infinitesimal
 CP map once the continuous-time limit is performed. This property is known as {CP-divisibility} and is regarded
 as an extended definition of quantum Markovianity \cite{reviewNMM}. Indeed, any CP-divisible dynamics can be shown to
 obey a ME where the Hamiltonian and dissipator are generally time dependent but, importantly, 
 the rate(s) appearing in the dissipator are guaranteed to be non-negative at any time. 
 
Therefore, a CM with step-dependent $\hat H_S$, $\eta$ and $\hat V_{Sn}$ will lead in the continuous-time limit to
a general ME of the form
 \begin{align}
\frac{{\rm d}\rho}{{\rm d} t}=-i \,[\hat H_S(t)+\hat H'_S(t),\rho]+ \,{\cal D}(t)[\rho]\label{ME-td}\,\,
\end{align}
with [\cf\eq\eqref{diss}] $\Gamma=\Gamma(t)\ge0$ at any $t$. Physically, 
this dynamics can still be considered to be essentially Markovian in that, during each infinitesimal time
interval ${\rm d}t$, there exists a Lindblad ME which describes it exactly. The crucial point is that,
no matter whether or not $\eta$ is the same for all the ancillas, these are initially uncorrelated.

\subsection{Initially correlated ancillas}\label{ICA}

All the above arguments, in particular \eqs\eqref{drho2}, \eqref{ME} and \eqref{ME-td}, do not hold
any more if the initial product bath state in \eq\eqref{sigma0} is replaced with a correlated one. In the latter case, due to the
pre-existing correlations between the bath ancillas, as soon as $S$ starts interacting with the bath it gets correlated with
the ancillas,  in general even those with which it still has to collide [see \fig1(b)]. This 
clearly endows the CM with memory in that past history affects the future dynamics. The open
dynamics of $S$ in general is no more described by a Lindblad ME, not even a time-dependent one as in \eq\eqref{ME-td}.
The reason is that, since $S$ and ancilla $B_n$ are no more in a product state before colliding with each other, the 
single-collision map on $S$ is no longer ensured to be completely positive as the one in \eq\eqref{eps}. 
Thereby, the resulting dynamical map will not be CP-divisible.

It is significant in this respect that CMs with initial correlated bath states can be constructed whose
corresponding dynamical map for $S$ reproduces indivisible quantum channels \cite{rybar2012,sabrinaCM}.

\subsection{Collision models versus standard system-bath models}

Technically, a CM is a microscopic system-bath model. This aspect is especially
useful in quantum thermodynamics applications, e.g. to connect the Landauer principle with
a microscopic framework \cite{lorenzo2015,pezzuto2016}, taking advantage from the intrinsic simplicity of CMs
which often allows for analytical calculations. That said, based on the definition 
reviewed in Section \ref{defCM}, note that in a CM the total Hamiltonian of $S$ and $B$ 
has the form
\begin{equation}
\hat H=\hat H_S+\sum_{n} f_n(t)\,\hat V_{Sn}
\end{equation}
with $f_n(t)$ equal to 1 during the time interval when the $n$th collision takes
place and zero otherwise. Hence,
the joint system-bath Hamiltonian is {\it time-dependent}. As anticipated 
in the Introduction, this is not what one usually expects
when dealing with an open dynamics in the presence of a reservoir (unless the dynamics
{\it per se}  is of a collisional nature (as \eg in \rref\cite{datta}).

\section{Emergence  of collision models  in a quantum optics setup}

As anticipated in the Introduction, we will now illustrate a quantum optics setup where a CM
can be naturally defined whenever the conditions for applying the input-output formalism are matched. We 
first define the Hamiltonian model and review the basics of input-output formalism.

\subsection{Input-output formalism}\label{io-theory}

Assume to have a generic system $S$ with free Hamiltonian $\hat H_S$ coupled to a continuum of bosonic modes (henceforth referred to as the ``field").
The free Hamiltonian of the field reads
\begin{equation}
\hat H_f=\int\! {\rm d}\omega\,\omega\,\hat a^\dag(\omega)a^\dag(\omega)\,,\label{Hf}
\end{equation}
where $\hat a(\omega)$ [$\hat a^\dag(\omega)$] annihilates (creates) a photon of frequency $\omega$ and with the integral
running over the entire real axis (similarly for all the integrals appearing henceforth). The field
operators obey the commutation relations $[\hat a(\omega), \hat a^\dag(\omega')]=\delta(\omega{-}\omega')$
and $[\hat a(\omega), \hat a(\omega')]=[\hat a^\dag (\omega), \hat a^\dag (\omega')]=0$.
The coupling between $S$ and the field is described by the interaction Hamiltonian 
\begin{equation}
\hat V=\int\! {\rm d}\omega\,\sqrt{\frac{\gamma }{2\pi}}\,\left( \hat b\, \hat a^\dag(\omega)+\hat b^\dag\, \hat a(\omega)\right) \,,\label{Vq0}
\end{equation}
where $\hat b$ and $\hat b^\dag$ are operators on $S$. Note that $S$
is coupled to all the field modes with the same strength. This is a key point, especially for establishing the connection with CMs.

In the interaction picture with respect to $\hat H_0=\hat H_f$, the joint state of $S$ and the field 
evolves as
\begin{equation}
\frac{{\rm d}\sigma}{{\rm d}t}=-i [\hat H_S+ \hat V(t),\sigma]\label{dsigma1}
\end{equation}
with 
\begin{equation}
\hat V(t)=\int\! {\rm d}\omega\,\sqrt{\frac{\gamma }{2\pi}}\,\left( \hat b\, \hat a^\dag(\omega)e^{i\omega t}+\hat b^\dag\, \hat a(\omega)e^{-i\omega t}\right)\,.\label{Vt}
\end{equation}
The form of \eq\eqref{Vt} suggests to define a time-dependent operator as \cite{gardiner}
\begin{equation}
\hat a_{\rm in}(t)=\frac{1}{\sqrt{2\pi}}\int\! {\rm d}\omega\,a(\omega)e^{-i\omega t}\,\label{input}
\end{equation}
called ``input operator" or ``quantum white noise operator" (one usually deals with an output operator as well \cite{gardiner}, but here it suffices to look only at 
the former since our focus is the open dynamics of $S$). The input operator can be viewed as the Fourier-transform of $\hat a(\omega)$ in the time domain.
The remarkable property of the input operator is that
\begin{equation}
[\hat a_{\rm in}(t), \hat a^\dag_{\rm{in}t}(t')]=\delta(t-t')\,,\label{comm}
\end{equation}
while of course $[\hat a_{\rm in}(t), \hat a_{\rm{in}t}(t')]=[\hat a^\dag_{\rm in}(t), \hat a^\dag_{\rm{in}t}(t')]=0$. 
Definition \eqref{input} allows to arrange the interaction Hamiltonian \eqref{Vt} as
\begin{equation}
\hat V(t)=\sqrt{\gamma}\left( \hat b\, \hat a_{\rm in}^\dag(t)+\hat b^\dag\, \hat a_{\rm in}(t)\right)\,. \label{Vt2}
\end{equation}
Even at this stage, the analogy with a CM should be evident: one can think of defining a (continuous) set of independent bosonic modes (input modes), 
labeled with $t$, whose respective ladder operators $\{\hat a_{\rm in}(t), \hat a^\dag_{\rm in}(t)\}$ commute at different times. As shown by \eq(\ref{Vt2}), $S$ interacts with
these modes in succession without ever interacting twice with the same mode. Moreover, the input modes are not mutually interacting 
since $\hat V(t)$ only couples $S$ to each of them. Therefore, apart from the continuous nature of the bosonic reservoir in the present model, we see
that the dynamics proceeds in analogy with a CM (see Section \ref{defCM}) with the input modes playing the role of bath ancillas. 
We show next how to construct a {\it discrete} CM that fully complies with the definition in Section \ref{defCM}.

\subsection{Time discretisation}\label{timedis}

We now discretise time in formal accordance with Section \ref{LME} [see \eq\eqref{step} and related text]. 
We thus split the overall time $t$ into shorter intervals, each of duration $\Delta t$, such that $t=N \Delta t$ 
with $\Delta t$ given by \eq\eqref{step} and with $t_n=n\Delta t$ (where $n=0,1,..,N$) becoming the discrete time variable. 

Through the Suzuki-Trotter formula \cite{suzuki} the evolution operator corresponding to \eq\eqref{dsigma1}
can be decomposed as
\begin{equation}
\hat U(t)=\lim\limits_{N\rightarrow\infty} \hat U_N\cdots \hat U_1\,\label{Ut2}
\end{equation}
with
\begin{eqnarray}
\hat U_n=e^{-i\int_{t_{n-1}}^{t_n} \!\!{\rm d}t'\,\left(\hat H_S{+}\hat V(t')\right)}\,,\label{Un}
\end{eqnarray}
for $n=1,2,...,N$.
We can now express the integral of $\hat V(t')$ (appearing in the exponent) in each time interval as
\begin{equation}
\int_{t_{n-1}}^{t_n} \!\!{\rm d}t'\,\hat V(t')=\sqrt{\gamma}\left( \hat b\, \hat \alpha_{n}^\dag+\hat b^\dag\, \hat \alpha_{n}\right)\!\sqrt{\Delta t}\,,\label{sqrt}
\end{equation}
where we defined the discrete set of operators
\begin{align}
{\hat \alpha}_n=\frac{1}{\sqrt{\Delta t}}\int_{t_{n-1}}^{t_n}\!\!{\rm d}t'  \,\,{\hat a}_{\rm in}(t')\label{alphan}\,,
\end{align}
which, due to \eq\eqref{comm}, fulfil the commutation rules $[\hat \alpha_n,\hat \alpha^\dag_m]=\delta_{nm}$ and 
$[\hat \alpha_n,\hat \alpha_m]=[\hat \alpha_n^\dag,\hat \alpha^\dag_m]=0$. 

The evolution operator in each interval now reads
\begin{equation}
\hat U_n=e^{-i \left[\hat H_S\Delta t+\sqrt{\gamma}\,\left( \hat b\, \hat \alpha_{n}^\dag+\hat b^\dag\, \hat \alpha_{n}\right)\sqrt{\Delta t}\right]}\,.
\end{equation}
The $\sqrt{\Delta t}$ in the above follows from the definition \eqref{alphan} where the factor 
$1/\sqrt{\Delta t}$ is required in order to ensure bosonic commutation rules of 
operators $\{\hat \alpha_n\}$. 
We illustrate next that in the CM picture the elementary evolution operator no longer features
$\sqrt{\Delta t}$ but only $\Delta t$ [in line with \eq\eqref{USn1}] with 
the system-ancilla coupling strength acquiring a dependence on $\sqrt{\Delta t}$.

\subsection{Collision model definition}\label{CMQ}

Indeed, if the right-hand side of \eq\eqref{sqrt} is multiplied and divided by ${\Delta t}$ then \eq\eqref{Un} can be
arranged as
\begin{eqnarray}
\hat U_n=e^{-i (\hat H_S + \hat V_{Sn})\Delta t}\label{Un3}\,,\label{Un3}
\end{eqnarray}
where we defined
\begin{equation}
\hat V_{Sn}=\frac{1}{\Delta t}\,\int_{t_{n-1}}^{t_n} {\rm d}t'\,\hat V(t')=\sqrt{\frac{\gamma}{\Delta t}}\,\left( \hat b\, \hat \alpha_{n}^\dag+\hat b^\dag\, \hat \alpha_{n}\right)\,.\label{VsnQ}
\end{equation}
By comparing \eqs\eqref{Un3} and \eqref{VsnQ} with \eqs\eqref{USn1} and \eqref{Hs}, we see that
we can indeed construct a CM where each discrete input mode defined by \eq\eqref{alphan} embodies a bath ancilla 
whose collision with $S$ is described by the interaction Hamiltonian $\hat V_{Sn}=g \,\hat v_{Sn}$ with
\begin{eqnarray}
g=\sqrt{\frac{\gamma}{\Delta t}}\,,\,\,\,\hat v_{Sn}=\hat b\, \hat \alpha_{n}^\dag+\hat b^\dag\, \hat \alpha_{n}\,.\label{param}
\end{eqnarray}
The system-ancilla coupling strength $g$ thus {\it diverges} in the limit $\Delta t\rightarrow 0$ as $1/\sqrt{\Delta t}$.

The CM defined this way belongs to the class of CMs where ancillas are not mutually interacting. Whether or not
the evolution of $S$ is described by the Lindblad ME (in the continuous-time limit) depends on the existence of
initial correlations between the ancillas. This in turn depends on the {\it field's initial state}. 

If the field is initially in a state such that the input
modes are in a product state, then $S$ is ensured to obey a ME of the form \eqref{ME} or the more general one \eqref{ME-td}.
In these cases, the resulting Lindblad ME is specified by $\hat H'_S$ and $\cal D$ defined by \eqs\eqref{HpS} and \eqref{diss} 
with $g$ and $\hat v_{Sn}$ given by \eq\eqref{param} and where the ancilla's state $\eta$ (this could be $N$-dependent as we will see)
is the state of each discrete input mode corresponding to the field's initial state. Importantly, the rate $\Gamma$ [\cf\eq\eqref{rate}] entering the dissipator \eqref{diss} here is given by
\begin{equation}
\Gamma= g^2 \Delta t= \gamma\,,\label{rateQ}
\end{equation}
and is thus ensured to remain {\it finite} in the continuous-time limit thanks to the aforementioned $g$'s divergence as $g\sim 1/\sqrt{\Delta t}$.

If the conditions in order for the Lindblad ME to hold are met, the specific form taken by the ME depends on the effective ancilla's state $\eta$, which indeed 
enters the definition of both $\hat H'_S$ and $\cal D$ [\cf\eqs\eqref{HpS}, \eqref{diss} and \eqref{jump}]. We illustrate this by considering 
next two paradigmatic initial states of the field which are expected to lead to the familiar spontaneous emission 
master equation and the optical Bloch equations, respectively.

\subsection{Vacuum state}

When the field is initially in the vacuum state each input mode defined by \eq\eqref{alphan} is correspondingly
in its own vacuum state $|0\rangle_n$ \cite{nota-vacuum}. Accordingly, in the CM picture, the bath is initially in a product state as 
in \eq\eqref{sigma0} with 
$$\eta=|0\rangle_n\langle 0|\,\,.$$
Then, from \eqs\eqref{HpS}, \eqref{diss}, \eqref{jump} and \eqref{param} it follows that 
\begin{align}
\hat H'_{S}&=g\, \,_n\!\langle 0|\hat v_{Sn}|0\rangle_n=0 \label{}\\
{\mathcal D}[\rho]&=\gamma \left(\hat{b}\,\rho\,\hat{b}^\dag\meno\tfrac{1}{2}[\hat b^\dag\hat b,\rho]_+\right)\,,\label{diss-vac}
\end{align}
where in the latter equation we used
\begin{equation}
_n\!\langle k|\hat v_{Sn}|0\rangle_n=\hat b\,\,_n\!\langle k|\hat \alpha_{n}^\dag|0\rangle_n=\delta_{k1}\,\hat b\label{id}
\end{equation}
with $\{|k\rangle_n\}$ (for $k=0,1,...$), denoting the Fock-state basis for the $n$th input mode. Note that in this specific case 
the bath-induced Hamiltonian $\hat H'_S$ does not arise.

Passing to the continuous-time limit, we end up with 
$$\dot{\rho}=-i[\hat H_S,\rho]+\gamma \left(\hat{b}\,\rho\,\hat{b}^\dag\meno\tfrac{1}{2}[\hat b^\dag\hat b,\rho]_+\right)\,,$$
which is, ax expected, the usual ME describing spontaneous emission (or loss).

\subsection{Coherent state}\label{coherent}

Consider next the case that the field is initially in a single-mode coherent state of frequency $\omega$ \cite{loudon}
\begin{equation}
|z\rangle{=}e\,^{\int \!\!{\rm d}\omega'\,\delta(\omega'-\omega)\left(z\hat a^\dag(\omega')-z^* \hat a(\omega')\right)}\,|0\rangle\,\label{coher}
\end{equation}
where $|z|^2$ is the average number of photons. By inverting \eq\eqref{input} we get
\begin{equation}
\hat a(\omega)=\frac{1}{\sqrt{2\pi}}\int\! {\rm d}t \,a_{\rm in}(t)e^{i\omega t}\,,\label{input2}
\end{equation}
hence
\begin{align}
\int \!\!{\rm d}\omega'\,\delta(\omega'-\omega)\hat a(\omega')&=\frac{1}{\sqrt{2\pi}}\int\! {\rm d}t \,a_{\rm in}(t)e^{i\omega t}\nonumber\\
&=\frac{1}{\sqrt{2\pi}}\sum_n\! \int_{t_{n-1}}^{t_n}\! {\rm d}t \,a_{\rm in}(t)e^{i\omega t}\,.\nonumber
\end{align}
For $\Delta t$ small enough, the exponential inside the last integral can be well-approximated by $e^{i\omega t_n}$. Thereby,
\begin{align}
\int \!\!{\rm d}\omega'\,\delta(\omega'-\omega)\hat a(\omega')\,\rightarrow\,&\frac{1}{\sqrt{2\pi}}\sum_n\!e^{i\omega t_{n}}\!\! \int_{t_{n-1}}^{t_n}\!\!\! {\rm d}t \,a_{\rm in}(t)=\nonumber\\
&=\frac{\sqrt{\Delta t}}{\sqrt{2\pi}}\sum_n e^{i\omega t_{n}}\hat \alpha_n\,,\label{eq-vac}
\end{align}
where we used \eq\eqref{alphan}\footnote{Operators \eqref{alphan} should be intended as defined for any $n$ running over all integers. Yet, only those for $0\le n \le N$ are involved in the dynamics occurring in the time interval $[0,t]$.}.
Correspondingly, in the same limit [\cf\eq\eqref{coher}]
\begin{equation}
|z\rangle\rightarrow \exp{\left[\sum_n \xi_n\,\hat \alpha^\dag_n-\sum_n\xi_n^* \hat \alpha_n\right]}|0\rangle\,,\label{coher2}
\end{equation}
where we set
\begin{equation}
\xi_n=\left(\frac{z}{\sqrt{2\pi}}\,e^{i\omega t_{n}}\right)\sqrt{\Delta t}\,\,.\label{xin}
\end{equation}
State \eqref{coher2} can be factorised as
\begin{equation}
|z_N\rangle=\bigotimes_{n} \,e^{\ \xi_n\hat \alpha^\dag_n-\xi_n^* \hat \alpha_n}\,|0\rangle_n=\bigotimes_{n}{\hat D}_n(\xi_n)|0\rangle_n\,,\label{zN}
\end{equation}
where $\hat D_n(w){=}e^{\,w\,\hat \alpha^\dag_n{-}w^* \hat \alpha_n}$ 
stands for the displacement operator on the $n$th input mode.

Based on \eqs\eqref{coher2} and \eqref{zN}, we see that once time has been discretised and 
the CM defined accordingly (see Subsections \ref{timedis} and \ref{CMQ}), the initial system-bath state 
reads
\begin{equation}
\sigma_{0}\ug\rho_{0}\otimes\left(\eta_1\otimes\eta_2\otimes...\otimes\eta_N\right)\,\label{sigma0bis}
\end{equation}
with each ancilla in the coherent state 
\begin{equation}
\eta_n=\hat D_n(\xi_n)\,|0\rangle_n\langle 0|\hat D_n^\dag(\xi_n)\,.\label{etan}
\end{equation}
At variance with \eq\eqref{sigma0}, here the ancillas are not in the 
same state. Still, the initial bath state is a {\it product} one which ensures that \eq\eqref{drho2} holds
in the present case as well (see Section \ref{TDLME}) with the Hamiltonian and dissipator given by [\cf\eqs\eqref{HpS}, \eqref{diss}, \eqref{jump} and \eqref{param}]
\begin{align}
\hat H'_{S}&=g\,(\xi_n^*\hat b +\xi_n\hat b^\dag)\,,\label{HpSc}\\
{\mathcal D}[\rho]&=\gamma \left(\hat{b}\,\rho\,\hat{b}^\dag\meno\tfrac{1}{2}[\hat b^\dag\hat b,\rho]_+\right)\nonumber\\
&\,\,\,\,\,\,+ \gamma \left(\hat{c}_n\,\rho\,\hat{c}_n^\dag\meno\tfrac{1}{2}[\hat c_n^\dag\hat c_n,\rho]_+\right)\,,\label{Dc}
\end{align}
where $\hat c_n$ is an operator on $S$ depending on $\xi_n$ defined by
\begin{equation}
\hat c_n=\xi_n^*\hat b+ \xi_n\hat b^\dag\,.\label{cn}
\end{equation}
\eq\eqref{Dc} is due to [\cf \eq\eqref{jump}]
\begin{align}
 _n\!\langle k |{\hat D}^\dag_n(\xi_n)\hat v_{Sn}{\hat D}_n(\xi_n)|0\rangle_n&=\, _n\!\langle k |\left(\hat v_{Sn}+\hat c_n\right)\!|0\rangle_n\nonumber\\
&=\delta_{k1}\,\hat b+\delta_{k0}\,\hat c_n\,\label{Lk}
\end{align}
with $\{|k\rangle_n\}$ again standing for the basis of Fock states of the $n$th input mode.
More details on the derivation of \eqs\eqref{HpSc}, \eqref{Dc} and \eqref{Lk} are given in the Appendix.

%&{=}\tfrac{\sqrt{\gamma}}{\sqrt{\Delta t}}\,(\xi_n^*\hat b {+}{\rm H.c.}){=}\sqrt{\tfrac{\gamma}{2\pi}}z\,(e^{-i\omega t_n}\hat b +e^{i\omega t_n}\hat b^\dag)\nonumber\\
In the continuous-time limit, $\Delta t\rightarrow 0$. Correspondingly, $g\rightarrow \infty$ and $\xi_n \rightarrow 0$ according to \eqs\eqref{param} and \eqref{xin}, respectively. As observed already, the limit does not affect the rate $\Gamma$ which stays finite
[see \eq\eqref{rateQ}]. Therefore, \eq\eqref{Dc} reduces to the same dissipator arising when the field is initially in the vacuum state [see \eq\eqref{diss-vac}] 
since the term featuring $\hat c_n$ [see \eq\eqref{cn}] vanishes due to $\xi_n\rightarrow 0$ for $N\rightarrow\infty$. Unlike the case in the previous subsection, however,
now the Hamiltonian term survives the continuous-time limit since, due to \eqs\eqref{param} and \eqref{xin},
$$g\,\xi_n=\sqrt{\frac{\gamma}{\Delta t}}\,\frac{z\sqrt{\Delta t}}{\sqrt{2\pi}}\,e^{i\omega t_{n}}=\sqrt{\frac{\gamma}{2\pi}}z\,e^{i\omega t_{n}}\rightarrow \sqrt{\frac{\gamma}{2\pi}}z\,e^{i\omega t}\,.$$
Accordingly, the ME governing the open dynamics of $S$ is obtained as
\begin{align}
\dot{\rho}=&-i[\hat H_S+\sqrt{\tfrac{\gamma}{2\pi}} \,|z|\,(\hat b e^{-i(\omega t+\phi) } +\hat b^\dag e^{i\left(\omega t+\phi\right)}),\rho]\,+\nonumber\\
&+ \gamma\left(\hat{b}\,\rho\,\hat{b}^\dag\meno\tfrac{1}{2}[\hat b^\dag\hat b,\rho]_+\right)\,,\label{bloch}
\end{align}
which corresponds, as expected, to the standard optical Bloch equations (here $\phi$ is such that $z=|z|e^{i\phi}$).

As anticipated previously, this instance in particular highlights that in the passage to the continuous-time limit one has to take into account that 
also the ancilla's state could depend on the time step $\Delta t$, as shown here by \eqs\eqref{xin} and \eqref{etan}.

It is worth pointing out that one could derive \eq\eqref{bloch} through a semiclassical approach by assuming a classical field quency $\omega$ driving $S$ 
and modifying accordingly the free Hamiltonian of $S$ which would thus become time-dependent. The bath ancillas would now be
initially prepared each in the vacuum state as in the previous section. Both the semiclassical approach and the fully quantum one that we followed above can be thus formulated 
in terms of corresponding CMs.

We also note that the fact that the system-ancilla coupling strength diverges as $\Delta t\rightarrow 0$ [\cf\eq\eqref{param}] ensures the validity of
condition \eqref{approx1} that is required in order for the low-order expansion \eqref{USn2} to hold.

\subsection{Occurrence of  correlated bath states}

The two instances of initial field states that we considered (vacuum and coherent state) both correspond, in the CM picture, to initially uncorrelated bath states. 
This agrees with the fact that in both situations the open dynamics of $S$ is well-known to be fully Markovian and described by a Lindblad ME. While other
instances of this kind can be made, it should be clear however that, in general, the initial bath state of the CM constructed in the way described above will be a 
{\it correlated} one. A simple example to see this is to consider a single-photon state
$$|\Psi\rangle=\int {\rm d}\omega\,\psi(\omega)\hat a^\dag(\omega) |0\rangle\,,$$
which in terms of input-mode operators \eqref{input} reads
\begin{equation}
|\Psi\rangle=\int\!{\rm d}t\,\left[\frac{1}{\sqrt{2\pi}}\int {\rm d}\omega\,\psi(\omega)e^{-i\omega t}\right]\hat a^\dag_{\rm in}(t) |0\rangle\label{ent}\,.
\end{equation}
%\int {\rm d}\omega\,\psi(\omega)\frac{1}{\sqrt{2\pi}}\int{\rm d}t\,\hat a^\dag_{\rm in}(t)e^{-i\omega t} |0\rangle=
Once time is discretised and the CM constructed, this state will generally give rise to a multipartite {\it entangled} state of the ancillas. The reasoning developed in Section \ref{defCM} to end up with a ME of the form 
\eqref{ME}, or even \eqref{ME-td} in a more general case, is no longer valid (see discussion in Subsections \ref{TDLME} and \ref{ICA}). We note that 
it was recently shown in terms of non-Markovianity measures \cite{reviewNMM} that the open dynamics of an atom undergoing scattering with a single-photon wavepacket in a linear-dispersion-law waveguide is generally NM \cite{duke}. It is also significant that NM CMs where the bath ancillas are initially in a ``single-photon" state were recently
studied and their strong NM nature stressed \cite{darius} (although the ancillas were modeled therein as qubits instead of harmonic oscillators).

\section{Physical origin of the  collision model}

Although a specific one, the quantum optics framework considered here allows to 
understand in more depth a distinctive feature of CMs. As mentioned in the Introduction, a CM enables the derivation 
of a Lindblad ME essentially without resorting to any approximations \cite{nota-barchielli}: 
one simply needs to pass to the continuous-time limit. 
This is a further remarkable difference from usual system-bath microscopic
models where instead working out Lindblad MEs demands to combine approximations, in particular the 
well-known Born-Markov approximation \cite{books}. These are typically associated with the 
shortness of the bath autocorrelation time compared to the characteristic time scale of the system-bath interaction.

In the present quantum optics scenario, the key properties enabling the construction of the CM are 
the assumption that the field spectrum is infinite alongside the flat coupling strength 
in the interaction Hamiltonian \eqref{Vq0} \cite{nota-spectrum}. The former allows to define independent input modes 
at different times [\cf\eq\eqref{input}], while the latter ensures that $S$ will interact with these one at a time. It is
as if $S$ keeps exploring and interacting with ``fresh" bath subunits, each subunit being totally unaware of previous interactions
of $S$ with other subunits. This would not be the case if the coupling strength in \eq\eqref{Vq0} were not flat: $S$ 
would interact with more than one subunit at once. The above should make clear that the
considered Hamiltonian model in fact guarantees a {\it zero} autocorrelation time of the bath and it is precisely this
property that enables to construct the CM. Once this is defined, the Lindblad ME then follows with no more
assumptions from the complete positivity of the collision map \eqref{eps}, which in the continuous-time
limit becomes infinitesimal (any infinitesimal completely positive dynamical map obeys a Lindblad ME \cite{books}). This
clarifies why the CM yields ME \eqref{ME-td} in the most general case if the ancillas are initially uncorrelated.

\section{Conclusions}

Quantum CMs embody an attractive theoretical tool that is becoming more and more used to investigate
quantum non-Markovianity within the general context of open quantum systems theory. Despite these advantages, 
some features in the abstract definition of a CM may raise concerns on a merely physical ground. 
We illustrated here how a class of open dynamics occurring in quantum optics can be effectively described from the 
viewpoint of a suitably defined CM, whose formulation is built upon the input-output formalism. In this 
well-defined physical scenario, typical CM issues, such as the time dependence of the system-bath Hamiltonian, 
the absence of inter-ancilla interactions, the possibility of initially correlated bath states and 
the subtleties in the passage to the continuous-time limit, appear natural and their physical
origin or interpretation clear.

Initial field's states yielding initially correlated ancillas is not the only way to introduce
a memory mechanism in the quantum optics CM considered here. Another way is 
to embed $S$ into a larger system, \eg one made out of $S$ and a ``memory" $M$ with 
only the latter one coupled to the input field. A further possibility is to impose geometrical
constraints on the bosonic field, such as adding a perfect mirror giving rise to a hard-wall
boundary condition. This introduces a feedback mechanism \cite{tufa1} that generally results in NM 
behaviour of $S$ \cite{tufa2}. One can show that this is equivalent to allowing $S$ to interact
with each discrete input mode {\it twice} (the time interval between the two interactions representing
the delay time), a dynamics that was tackled in \rref\cite{grimsmo2015} through a nice diagrammatic 
method.

Before concluding, we make some comments.

Given the bosonic nature of the field addressed here, which has a non-marginal role in the formulation 
of the input-output formalism, it is natural to ask if a CM can be constructed likewise from a fermionic field. 
Although non-trivial, the formulation of a fermionic input-output theory is possible as shown by
Gardiner in 2004 \cite{gardiner-f}, this making plausible the possibility to define a CM in the fermionic case as well.

In the CM considered here which we used to work out quantum optics MEs, each ancilla is a quantum harmonic oscillator.
It is worth observing that, as recently shown in \rref\cite{milburn2017}, quantum optics MEs can be derived as well through
a suitably defined discrete dynamics where the system interacts in succession with qubits.

We hope this work could help establish a new link between the CMs and quantum optics research areas.  
The former could draw inspiration from the framework addressed here as a basis for future developments 
in quantum non-Markovianity.
The latter could benefit from the advancements in the knowledge of NM dynamics that are being made through 
memory-endowed versions of CMs. For instance, it is interesting to explore whether recently-discovered 
NM Gaussian MEs \cite{gaussian} can be somehow connected to CMs.

\paragraph*{Acknowledgments}

Fruitful discussions with A. Grimsmo, M. G. Genoni, A. Carollo, V. Giovannetti, G. M. Palma and S. Lorenzo are gratefully acknowledged.

\section*{Appendix}

\eqs\eqref{HpSc}, \eqref{Dc} and \eqref{Lk} are worked out through standard quantum optics calculations.
The displacement operator $\hat D^\dag(\xi_n)$ defines a unitary transformation that turns the annihilation operator 
$\hat \alpha_n$ into
\begin{align}
\hat D^\dag(\xi_n)\,\hat \alpha_n\,\hat D(\xi_n)&=\hat \alpha_n+\xi_n\label{unitary1}\,.
%\\\hat D(\xi_n)\,\hat \alpha_n\,\hat D^\dag(\xi_n)&=\hat \alpha_n-\xi_n^*\label{unitary2}\,.
\end{align}
We also note that in the case of \eq\eqref{etan}, the basis in the single-ancilla Hilbert space entering \eqs\eqref{diss} and \eqref{jump} is given by $\hat D(\xi_n)|k\rangle_n$ 
(with $\{|k\rangle_n\}$ the number-state basis) with all the $p_j$'s vanishing but the one corresponding to $\hat D(\xi_n)|0\rangle_n$.
Using \eq\eqref{unitary1} and recalling \eq\eqref{param} we get
\begin{align}
\hat H'_{S}&=g\,\, _n\!\langle 0|\hat D_n^\dag(\xi_n)\,\hat v_{Sn}\,\hat D_n(\xi_n)\,|0\rangle_n\label{}\nonumber\\
&=g \left(\hat b^\dag\, _n\!\langle 0|\left(\hat \alpha_{n}{+}\xi_n\right) |0\rangle_n+{\rm H.c.}\right){=}g\,(\xi_n^*\hat b +\xi_n\hat b^\dag)\,.\nonumber
\end{align}
Using \eq\eqref{unitary1} and its adjoint we get that $\hat v_{Sn}$ transforms as
\begin{align}
\hat D^\dag(\xi_n)\,\hat v_{Sn}\,\hat D(\xi_n){=}\hat v_{Sn}+(\xi_n^*\hat b+ \xi_n\hat b^\dag){=}\hat v_{Sn}+\hat c_n\nonumber\,,
\end{align}
hence \eq\eqref{Lk} holds.


\begin{thebibliography}{99}
\bibitem{nota-int} In the latter case, $\hat V_{Sn}$ is required to remain time-independent in the rotating frame. The extension
to the case that $\hat V_{Sn}$ is time-dependent is yet straightforward.
\bibitem{nota-vacuum}  The vacuum state of the $n$th input mode is defined by $\hat \alpha_n |0\rangle_n{=}0$. Based on \eq(\ref{eq-vac}), for $N{\rightarrow}\infty$, 
$\hat a(\omega){\rightarrow}\frac{\sqrt{\Delta t}}{\sqrt{2\pi}}\sum_n e^{i\omega t_{n}}\hat \alpha_n$. Correspondingly, in the same limit, $\hat a(\omega)\bigotimes_{n} \!|0\rangle_n{=}0$ \cite{mandel}, 
hence $|0\rangle{=}\bigotimes_{n} \!|0\rangle_n$.
\end{thebibliography}
\end{document}